# Primary and Secondary Hydration Forces between Interdigitated Membranes Composed of Bolaform Microbial Glucolipids

Niki Baccile,[a*] Viviana Cristiglio[b]

[a] Sorbonne Université, Centre National de la Recherche Scientifique, Laboratoire de Chimie de la Matière Condensée de Paris, LCMCP, F-75005 Paris, France

[b] Institut Laue-Langevin, 71 Avenue des Martyrs, 38042 Grenoble Cedex 9, France

**Abstract**

To better understand lipid membranes in living organisms, the study of intermolecular forces using the osmotic pressure technique applied to model lipid membranes has constituted the ground knowledge in the field of biophysics since four decades. However, the study of intermolecular forces in lipid systems other than phospholipids, like glycolipids, has gained a certain interest only recently. Even in this case, the work generally focuses on the study of membrane glycolipids, but little is known on new forms of non-membrane functional compounds, like microbial bolaform glycolipids. This works explores, through the osmotic stress method involving an adiabatic humidity chamber coupled to neutron diffraction, the short-range (< 2 nm) intermolecular forces of membranes entirely composed of interdigitated glucolipids. Experiments are performed at pH 6, when the glucolipid is partially negatively charged and for which we explore the effect of low (16 mM) and high (100 mM) ionic strength. We find that this system is characterized by primary and secondary hydration regimes, respectively insensitive and sensitive to ionic strength and with typical decay lengths of $\lambda_{H1}$= 0.37 ± 0.12 nm and $\lambda_{H2}$=1.97 ± 0.78 nm.

**Introduction**

Understanding the physical properties of biological membranes has long been a goal in biophysics and colloids science.[1–3] Due to their complex composition and the evident difficulties to study them *in-vivo*,[1] research is generally focused on the simplification of complexity by studying intermolecular forces in model lipid systems, and exploring both structural (electrical charges, bilayer flexibility, polar headgroup composition) and physicochemical (ionic strength, pH, temperature) parameters in both neutral and charged but also mixed compositions of neutral, charged phospholipids.[1,4–11]

Interactions in lipid lamellar phases are governed by a balance between attractive and repulsive forces.[12] Van der Waals attraction is counterbalanced by short- (< ~ 2 nm)[13] and long-range (> ~2 nm) repulsive forces,[5,14,15] where steric and hydration forces are typical short-range interactions while electrostatic and entropic (undulation) forces are most common at longer distances. To this regard, osmotic stress experiments are typically employed to obtain pressure-distance profiles,[7,9,16–20] which can be faced to both classical DLVO theory and its deviations describing intermolecular forces and colloidal interactions,[5,11,13,21] thus contributing to better understand model lipid membrane systems in relationship to living organisms.[5,10,13,19,22,23]

More recently, the quest of model lipid systems has drawn its attention away from classical phospholipids in profit of lipids characterized by a glycosylated headgroup, whereas glycolipids are minor but important components of biological membranes.[4,20,24] To this regard, the understanding of molecular interactions in glycolipid membranes is still in its infancy, because of the interesting hydration properties of sugars[25] and the broad variety of glycosidic headgroups. Stepping out of model lipid systems, a new class of entirely biobased compounds produced by microbial fermentation and characterized by a sugar headgroup, an aliphatic chain and a carboxylic acid end-group is gaining a large interest for its biobased origin, low cytotoxicity and potential applications as green surfactants.[26,27] These bolaform microbial glycolipids have an unpredictable, although rich, phase diagram, characterized by the molecular sensitivity to pH, which controls the carboxylic/carboxylate, $COOH/COO^-$, ratio and, consequently, the electrostatic interactions.[28–30] In a recent series of works, we have shown the ability of a single glucose bolaform lipid to form membranes in water at pH below 7 and composed of an interdigitated lipid structure.[30–32]

This work aims at studying, for the first time, the short-range molecular interactions of a bolaform glycolipid obtained by microbial fermentation, belonging to the family of biosurfactants and characterized by a single glucose moiety, a C18:0 chain and an end COOH group (GC18:0, Figure 1). This compound is known to self-assemble at acidic pH into an interdigitated lipid P$_{\beta,i}$ lamellar phase forming highly viscous/hydrogel solutions in water at concentrations above 1 wt% and T < 30°C.[32] Under typical conditions in bulk (C = 1-5 wt%, pH = 6-7, [NaCl] = 10-100 mM), the lamellar period at room temperature varies between 25 and 15 nm.[32]

We employ the osmotic stress technique inside an adiabatic humidity chamber[8,33] to draw pressure-distance profiles in the interlamellar distance range below 2 nm, where steric and hydration forces are predominant. An adiabatic humidity chamber provides an environment where the interlamellar $d$-spacing, $d_{(100)}$, can be controlled through relative humidity inside the chamber and easily adaptable to probe the interlamellar distance by using X-ray or neutron diffraction, the latter employed in this work. At high $RH\%$, the lamellar phase is hydrated and the thickness of the water layer increases, generally above 2 nm, after which long-range forces, like electrostatic repulsion, overwhelm Van der Waals attraction. At low $RH\%$, the interlamellar thickness decreases as a result of dehydration, and Van der Waals attraction overwhelms electrostatic repulsion, pushing the lamellae together. Below 1 nm, short-range repulsive interactions in lamellar systems generally contain steric and hydration components counterbalancing the Van der Waals forces. Establishing a pressure-distance relationship, $\Pi(d_w)$, with $\Pi$ being the osmotic pressure and $d_w$ the interlamellar water thickness, we will determine the nature, strength and decay length of the short ranges forces.

**Materials and methods**

*Products*. Acidic deacetylated C18:0 glucolipids (GC18:0) have been used from previously existing batch samples, the preparation and characterization ($^1$H NMR, HPLC) of which is published elsewhere.[31] Acid (HCl 37%), base (NaOH) and NaCl are purchased at Aldrich. MilliQ-quality water has been employed throughout the experimental process.

*Preparation of hydrogels*. Protocol of preparation and characterization of the lamellar phase from GC18:0 are reported elsewhere[32] and are adapted to this work. GC18:0 sample is dispersed in water, followed by sonication and adjustment of pH to the desired value and ionic strength. We prepare two solutions of C= 1 wt% in D$_2$O at pH = 6.2 and at [NaCl] = 16 mM and 100 mM. The pH is adjusted by using 1-5 µL of NaOH 1 M (0.1 M can also be used for refinement). The mixture is then sonicated between 15 and 20 min in a classical sonicating bath to reduce the size of the aggregated powder and until obtaining a homogenous, viscous, dispersion. To this solution, the desired volume of NaCl is added so to obtain a given total [Na$^+$] (= [NaOH] +

[NaCl]) molar concentration. To keep the dilution factor negligible, we have used a 5 M concentrated solution of NaCl. The mixture is then sonicated again during 15 min to 20 min and eventually vortexed two or three times during 15 s each. The solution can then be left at rest during 15 min to 30 min. The solution is highly viscous and it forms a gel at rest and it presents shear-thinning properties. The lamellar phase is characterized with neutron scattering[32] before depositing on a substrate for the adiabatic desiccation experiments.

*Adiabatic desiccation experiments using a humidity chamber*. The GC18:0 solutions are dispersed on two separate 5 cm x 2 cm silicon wafers by simple drop cast (volume dropped: 500 μL). To enhance homogeneous spreading of the solution onto the substrate, we have used a horizontal support levelled with a 2D spirit level. The silicon substrates are let drying in an oven at 40°C until a homogeneous coating is obtained. The samples are then introduced inside the humidity chamber,[34] provided at the D16 beamline at ILL (refer below for more details), and set under vacuum at T= 25°C. The temperature of the $D_2O$ water bath below the sample is modified to set the chamber at the desired *RH*% value. The humidity chamber is conceived to provide values of RH% with an error of ±0.01% RH. Technical details of the humidity chamber can be found in ref. [34] The sample at [NaCl]= 16 mM is let equilibrating at 98 *RH*% before studying, where relative humidity is lowered. The sample at [NaCl]= 100 mM sample is let equilibrating at 10 *RH*% and humidity is then increased.

*Neutron diffraction*: neutron diffraction experiments are carried out as described in ref. 47 on the D16 instrument at the Institut Laue-Langevin (ILL; Grenoble, France), using a wavelength $\lambda$= 4.5 Å ($\Delta\lambda/\lambda$= 0.01) and a sample-to-detector distance of 900 mm.[35] The focusing option provided by the vertically focusing graphite monochromator is used to maximize the incident neutron flux at the sample. The intensity of the diffracted beam is recorded by the millimeter-resolution large-area neutron detector (MILAND) $^3$He position-sensitive detector, which consists of 320 × 320 *xy* channels with a resolution of 1 × 1 $mm^2$. The samples are held vertically in a dedicated temperature-controlled humidity chamber and aligned on a manual 4-axis goniometer head (Huber, Rimsting, Germany) embedded in the humidity chamber. The chamber is mounted on the sample rotation stage, where the lipid multilayer stacks are scanned by rocking the wafers horizontally. Diffraction data are collected at a detector angle 2 θ of 12˚, by scanning the sample angle ω in the range -1 to 8˚, with a step of 0.05˚. Data analysis is performed using the ILL in-house LAMP software (www.ill.eu/instruments-support/computing-for-science/cs-software/all-software/lamp).[36] The classical *I* vs 2*θ* profile for each *RH*% is obtained by summarizing each integrated 2D image measured at a given value of ω. The lamellar spacing $d_{(100)}$ is obtained by a fitting the (100) peak position with a Gaussian profile. Intensities on the detector surface are corrected for solid angle and pixel efficiency by normalization to the flat incoherent signal of a 1 mm water cell.

The sample temperature in the chamber is maintained at 25°C during the measurements, and the humidity is varied by changing the temperature of the liquid reservoir generating the water vapor from 10°C to 24°C, leading to relative humidities ranging from to 10% to 98%. Each sample is investigated by increasing the humidity step by step without opening the chamber at any time during the humidity scan. After each change in relative humidity, the sample is equilibrated between 30 min to 2 h, where equilibration is followed through the evolution of the (100) diffraction peak position over time. Equilibration time is followed (by collecting ω-2θ scans) until the diffraction peak position reach a plateau. After equilibration, the rocking curve (ω scan between -1° and 8° with 0.05°) is recorded.

**Results and discussion**

The functional glucolipid GC18:0 (Figure 1) is obtained by hydrogenation[31] of the monounsaturated GC18:1 compound, produced by fermentation of glucose and fatty acids by the yeast *S. bombicola ΔugtB1*.[37,38] The phase behaviour of this compound in water below concentrations of 10 wt% depends on pH and it was shown that it undergoes a micellar-to-lamellar transition at room temperature when pH is decreased from 10 to 5 (Figure 1).[30–32] We have previously shown by small angle X-ray scattering that the membranes at acidic pH are composed of interdigitated GC18:0 molecules containing a mixture of COOH and COO$^-$ groups. The membrane has an overall thickness (hydrophilic and hydrophobic layers, respectively of thickness $T_h$ and Length, *L*, Figure 1) of about 3.6 nm,[30–32] where an error of about ±10% should be considered due to the fitting procedure. GC18:0 solutions at pH between 6 and 7 and ionic strength between 10 and 100 mM are highly viscous, possibly gels, with shear-thinning properties.[32] A typical GC18:0 bulk solution at concentration of 1 wt% is used in this study.

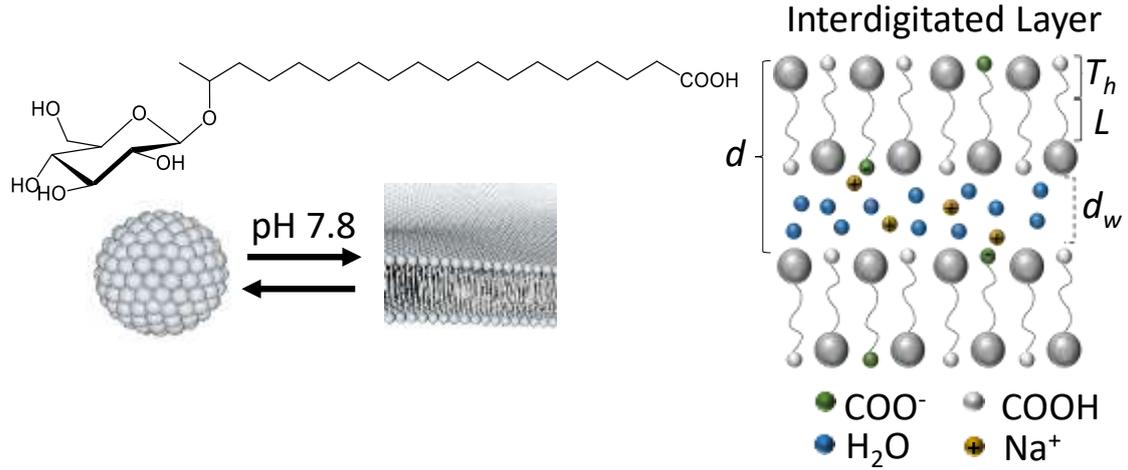

**Figure 1** - Molecular structure of bolaform glucolipid GC18:0. Headgroup is composed of *D*-glucose and the backbone is composed of a C18:0 fatty acid, with a free COOH group at the opposite end of the glucose moiety. GC18:0 is obtained by hydrogenation of the corresponding GC18:1 compound, produced by fermentation of sugar and vegetable oil with the yeast *S. bombicola ΔugtB1*. The GC18:0 lipid undergoes a micelles-to-lamellar phase transition and the latter is formed of interdigitated GC18:0 molecules. In the structure of the membrane, $T_h$ and $L$ are, respectively, the thickness of the hydrophilic and hydrophonic layers, $d_w$ is the thickness of the interlamellar water layer and $d$ is the lamellar period.

The GC18:0 solution, prepared in D$_2$O to enhance the contrast with neutrons, is drop-cast and allowed to dry on a silicon wafer, while the lamellar $d$-spacing is probed using neutron diffraction in a $\theta$-$2\theta$ configuration, with the relative humidity ($RH\%$) varying between 98% and 10% (Figure 2a). The repeating lamellar period, $d_{(100)}$, is traced against relative humidity, $d_{(100)}(RH\%)$ (Figure 2b), and eventually converted into a $\Pi(d_{(100)})$ relationship (Figure 3) using the following expression equalizing pressure and $RH\%$,[39,40]

$$\Pi = -N_A \left(\frac{k_b T}{V_w}\right) ln\left(\frac{RH\%}{100}\right) \qquad \text{Eq. 1}$$

with $\Pi$ being the osmotic pressure, $N_A$ the Avogadro contant, $k_b$ the Boltzmann's constant, $T$ the temperature in degrees Kelvin, $V_w$ the water molar volume and $RH\%$ the relative humidity. The thickness of the interlamellar water layer, $d_w$, is commonly obtained by subtracting the membrane thickness from $d_{(100)}$. Under high humidity conditions, above 80%, the (100) reflection settles at about $2\theta= 4°$, corresponding to $d_{(100)}$ of about 6 nm, while below 50%, the (100) reflection shifts towards $2\theta= 6°$, corresponding to $d_{(100)}$ between 4.1 and 4.5 nm. The $d_{(100)}(RH\%)$ profiles in Figure 2b show that salt has no influence on $d$-spacing at relative humidity below 40%. On the contrary, an important mismatch in $d$-spacing values between the 16 mM and 100 mM systems occurs above $RH\%= 40\%$, where $d_{(100)}$ is larger at lower salt concentration. These data confirm the trend observed in bulk for the same material by mean of neutron scattering,[32] and where the $d$-spacing was found to vary from 22 nm to 10 nm when salt concentration increases from 50 mM to about 300 mM. Similar trends were also found for other lipid lamellar phases when increasing ionic strength.[41,42]

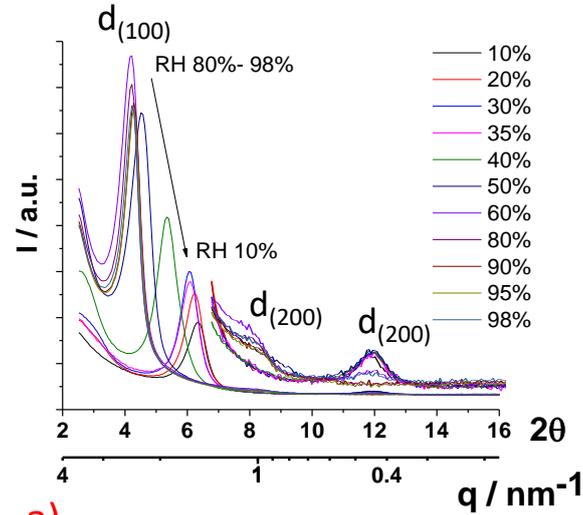

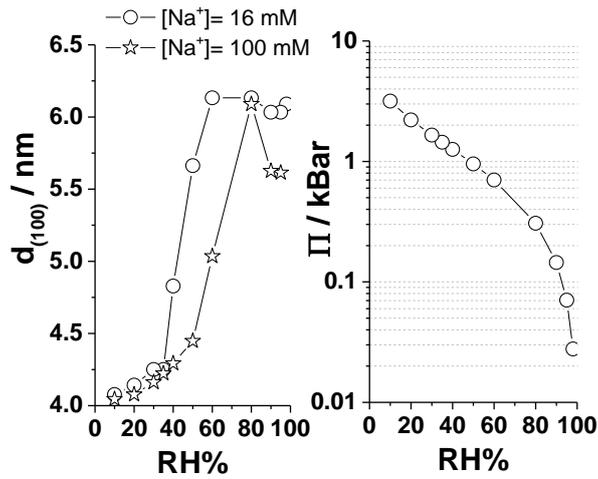

**Figure 2 -** a) **Evolution of the neutron diffraction patterns as a function of relative humidity, $RH\%$, measured on a GC18:0 solution (bulk data: C= 1 wt%, pH 6.2 ± 0.3, [NaCl]= 16 mM) drop-cast on a silicon (111) substrate. b) Evolution of the $d$-spacing with $RH\%$, and plot of the corresponding $\Pi(RH\%)$ relationship with $N_A$ being the Avogadro constant, $K_b$ the Boltzmann constant, $T$ the temperature in Kelvin degrees and $V_m$ the water molar volume.**

In order to establish a classical pressure-distance, $\Pi(d_w)$, relationship, $d_{(100)}$ must be converted into the thickness of the water layer between the lamellae, $d_w$, by mean of Eq. 2, where 3.6 nm is the thickness of the interdigitated layer of GC18:0 measured by SAXS at concentrations below 10 wt%. If this specific value was previously measured on the same material,[30–32] one should consider at least a ±10% error, which can lead to a significant uncertainty in the determination of $d_w$, especially at $d_w$< 0.5 nm To avoid this source of error, we will also consider $\Pi(d_{(100)})$ profiles in the quantification of intermolecular interactions, as discussed later.

The interaction terms contained in the expression of $\Pi(d_w)$, also known as the equation of state of the lamellar system, are shown in Eq. 3. $\Pi(d_w)$ contains both attractive (Van der Waals, *VdW*) and repulsive (steric, hydration, electrostatic, entropic, respectively *St*, *Hyd*, *El* and *Entr*) contributions.

$$d_w(RH\%) = d_{(100)}(RH\%) - 3.6\ nm \qquad \text{Eq. 2}$$

$$\Pi(d_w) = \Pi_{VdW} + \Pi_{St} + \Pi_{Hyd} + \Pi_{El} + \Pi_{Entr} \qquad \text{Eq. 3}$$

The lin-lin plot (Figure 3a) of the $\Pi(d_w)$ curves suggests a double exponential decay, confirmed by the log-lin plots in Figure 3b-d and where the frontier between the two regimes is at 4.2 nm < $d_{(100)}$ < 4.5 nm (0.5 nm < $d_w$ < 0.7 nm). The pressure below which the interlamellar distance is constant is generally referred to the as the disjoining pressure, it is commonly observed in osmotic stress experiments for water thicknesses above 2-3 nm and it can be described by the necessary force to overcome hydration forces.[16,21,43] In the present system, the disjoining pressure is set at about 1 kbar and it is identified by the grey symbols between 2 and 2.5 nm in Figure 3b-d. Hydration forces are generally found at interlamellar distances below 1 nm and they are characterized by a single exponential decay with a decay length, *λ*, between 0.2 – 0.4 nm.[9,11,43] In the same range of $d_w$, one can measure repulsive steric forces, corresponding to excluded volume steric interactions between polar groups, and with characteristic decay lengths smaller than 0.2 nm.[16] A double exponential fit of the $\Pi(d_{(100)})$ curves in Figure 3 yields two values of *λ*, *λ₁* ~ 0.3 nm and *λ₂* ~ 2 nm. If *λ₁* is compatible with the classical values of hydration decay lengths found in lamellar phases composed of surfactants or phospholipids,[9,11,44] *λ₂* is excessively larger and cannot be explained with classical short-range repulsion forces (steric and hydration). At the same time, the pressure range of *Π* = 1 ± 0.5 kbar reached between 1 and 2 nm is also excessively high for classical long-range forces such as electrostatic or entropic.[33,45] Tentative calculations of $\Pi_{El}(d_w)$ for $d_w$ above 0.7 nm and for any pressure regime identified in ref. [5] yield values below 1 bar, that is three orders of magnitude smaller than what we experimentally measure here. Pressure values of comparable magnitude are obtained for $\Pi_{Entr}$ calculated using the classical Helfrich formula[15,46] using typical bending modulus values in the order of 10-20 $k_B$T. In both cases, the calculated values for the pressure for $d_w > 0.7$ nm are at least 2 orders of magnitude smaller than what we find experimentally in Figure 3d, as also shown in [47].

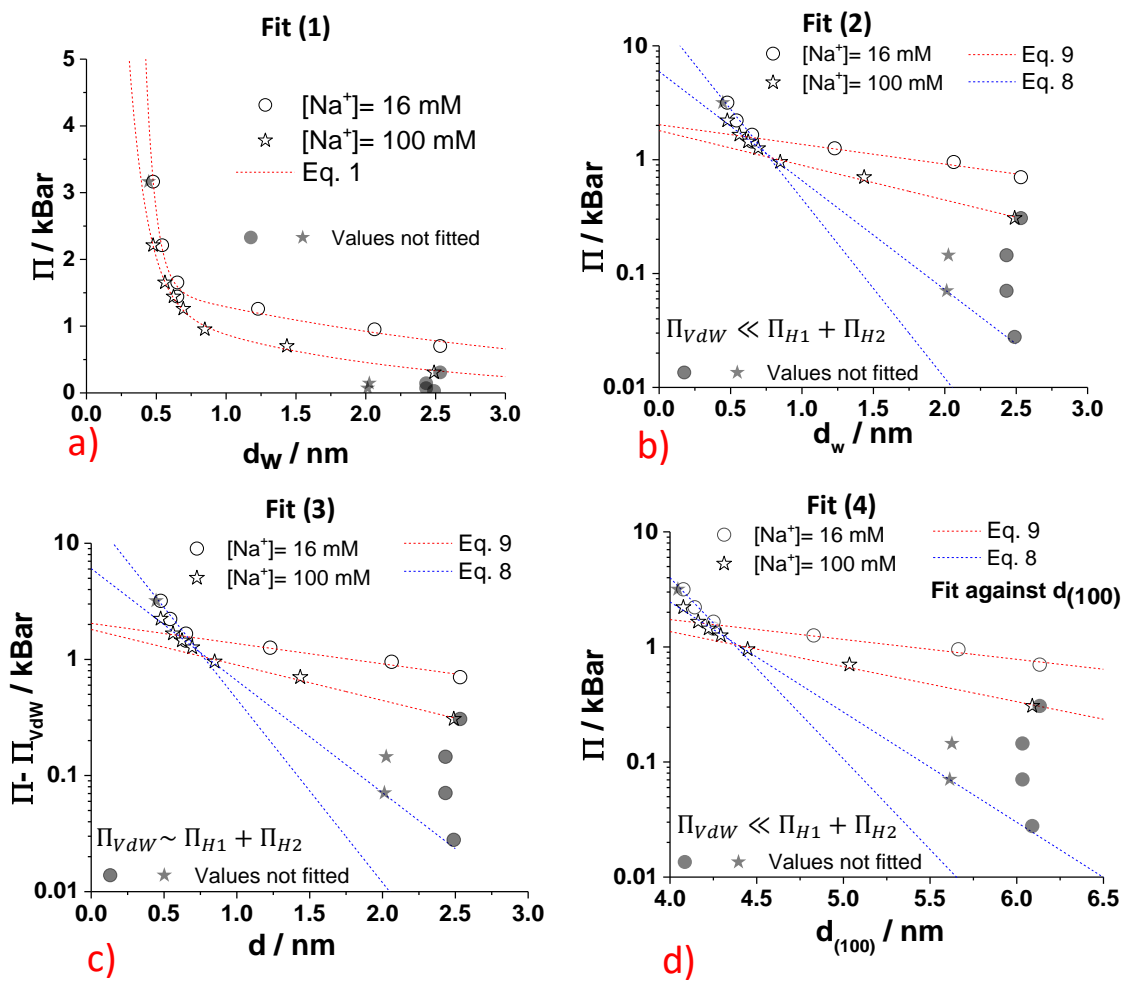

**Figure 3** - **Various representations of the pressure-distance plots derived from the osmotic stress experiments in Figure 2a. Fitting strategies (1) through (4), described in the main text, are respectively used to fit data in panels a) – d). Grey-filled values identify the disjoining pressure and they are not included in the fits.**

The considerations above show that if the $\Pi_{St}, \Pi_{El}, \Pi_{Entr}$ terms of $\Pi(d_w)$ can be neglected against $\Pi_{Hyd}$, one term accounting for the pressures at $d_w > 0.7$ nm is most likely missing, and its nature is exponential. Described long ago,[10,22,48] secondary hydration forces are commonly observed in charged lamellar systems in the presence of an electrolyte, and the typical reported $\lambda$ is contained between 1 and 3 nm,[10,13,21,22,43,47,48] values which are in perfect agreement with $\lambda_2$ measured in this system. Long-range hydration forces are commonly accepted as deriving from short-range repulsion due to ion exclusion from surface hydration layers, and longer range repulsion, arising from ionic dispersion interactions. In all cases, they have been clearly identified in lamellar systems at ionic strength from the millimolar to the molar range. In the systems in Figure 3, we study a low and high salt concentration, corresponding to [NaCl]= 16 mM and 100 mM in bulk gel, where $d_{(100)}$ is ~20 nm.[32] Preparing the sample for the pressure-distance experiments involves a deposition and drying step of the gel onto a silicon wafer, before introduction of the latter in the humidity chamber. After drying, $d_{(100)}$ ~4 nm, that is a shrinking factor of five of the $d$-spacing, and corresponding to a five-fold increase in the interlamellar NaCl concentration. In simplistic hypothesis that the interlamellar NaCl concentration is the same as the bulk NaCl concentration, one estimates the interlamellar NaCl concentration after drying to be set between 80 mM and 500 mM, respectively for the low and high ionic salt samples. These values are high enough to expect secondary hydration forces according to literature.[21,43,47] Under these circumstances, Eq. 3 can be simplified to Eq. 4, where the steric, electrostatic and entropic terms can be neglected while a second hydration term is introduced.

$$\Pi(d_w) = \Pi_{VdW} + \Pi_{Hyd1} + \Pi_{Hyd2} \qquad \text{Eq. 4}$$

Pressure-distance plots have been analyzed using four different fitting strategies derived from Eq. 4 and all classically employed in the literature reporting the study of intermolecular forces on similar systems. The discussion below presents each fitting strategy separately, giving the advantages and disadvantages, and establishing a range of values for the strength and length on the hydration interactions.

*Fit (1)*. It uses equation Eq. 4 to fit the $\Pi(d_w)$, where the expressions of $\Pi_{VdW}$ is given in Eq. 5, while the primary and secondary hydration components, $\Pi_{Hyd1}$ and $\Pi_{Hyd2}$, of the hydration pressure (Eq. 6, linearized in Eq. 7) are given in Eq. 8 and Eq. 9, respectively. This approach is classically employed by many authors in fitting lin-lin plots of pressure-distance profiles.[18,40] Fit (1) is the most rigorous approach, but, in order to reduce the number of free parameters in the fit to only four ($\Pi_{H1}$, $\Pi_{H2}$, $\lambda_{H1}$, $\lambda_{H2}$), it supposes to calculate the $\Pi_{VdW}$ term. To do so, one must calculate the Hamaker constant, $H$, but also have a good estimation of $T_h$ and $L$, respectively the thickness of the hydrophilic and length of the hydrophobic regions of the membrane, and to assume that the value of 3.6 nm, used to calculate $d_w$, for the bilayer thickness is also a good estimation. At room temperature, these parameters can either be calculated or measured. The Hamaker constant was calculated for a generic lipid bilayer to be $H = 5.1 \cdot 10^{-21}$ J at room temperature,[49] the structural parameters of the GC18:0 interdigitated layer were estimated from the fit of SAXS data[30–32] and are assumed here to be $T_h = 1.4$ nm, $L = 0.8$ nm and the total thickness, $(2T_h+L) = 3.6$ nm.

$$\Pi_{VdW} = \frac{H}{6\pi}\left(\frac{1}{d_w^3} - \frac{2}{(d_w + 2T_h + L)^3} + \frac{1}{(d_w + 2(T_h + L))^3}\right) \quad \text{Eq. 5}$$

$$\Pi(d_w)_{Hyd} = \Pi_H e^{-\frac{d_w}{\lambda_H}} \quad \text{Eq. 6}$$

$$Log(\Pi_{Hyd}) = Log(\Pi_H) - \frac{0.434}{\lambda_H} d_w \quad \text{Eq. 7}$$

$$\Pi(d_w)_{Hyd1} = \Pi_{H1} e^{-\frac{d_w}{\lambda_{H1}}} \quad ; d_w < 0.74 \pm 0.11 \ nm \quad \text{Eq. 8}$$

$$\Pi(d_w)_{Hyd2} = \Pi_{H2} e^{-\frac{d_w}{\lambda_{H2}}} \quad ; d_w > 0.74 \pm 0.11 \ nm \quad \text{Eq. 9}$$

Fit (2). In fit (2), we make the hypothesis according to which the contribution of the Van der Waals term is negligible across the entire $d_w$ range against the hydration terms, that is $\Pi_{VdW} \ll \Pi_{Hyd1} + \Pi_{Hyd2}$ in Eq. 4. This hypothesis holds for a system that does not follow the DLVO theory at small thickness of water layer, as this seems to be the case for lamellar lipid phases dominated by two hydration regimes.[47] Under this hypothesis, one can represent the pressure-distance curves in a log-lin plot (Figure 3b), as largely shown by other authors,[4,24] and in particular the hydration component (Eq. 6) can be linearized into Eq. 7. If the two hydration regimes are distinct enough, one can independently fit the short- and long-distance domains of the pressure-distance curves with equations Eq. 8 and Eq. 9 and extract $\Pi_{H1}$, $\Pi_{H2}$, $\lambda_{H1}$ and $\lambda_{H2}$, as this was classically done in lamellar systems governed by two hydration regimes.[10,13,22]

Fit (3). In fit (3) we employ exactly the same approach as in fit (2), but the Van der Waals contribution is not neglected anymore: $\Pi_{VdW}$ is calculated exactly as in fit (1) and subtracted to $\Pi(d_w)$, as proposed long time ago by Pashley and Israelachvili.[10,13] The resulting term is plot against $d_w$ in a log-lin scale (Figure 3c) and $\Pi_{H1}$, $\Pi_{H2}$, $\lambda_{H1}$ and $\lambda_{H2}$ terms are extracted from linear fits according to equations Eq. 8 and Eq. 9. This approach, to which the attractive DLVO contribution is accurately subtracted, was classically used by Pashley in the early studies of the double hydration regime.[10,13,22]

Fit (4). The drawback of fits (1)-(3) is the plot of pressure against $d_w$, being calculated using equation Eq. 2 and supposing a good estimate for the membrane thickness. We use the value of 3.6 nm determined by modelling SAXS profiles in bulk, but it is well-known that fitting of SAXS curves generally requires more than one free variable and acceptable fitting can occur with a broad set of numerical solutions. Although we believe that a membrane thickness of 3.6 nm is the best estimate, one must consider an error of least ± 10%, which may have a strong impact on the pressure-distance profiles at low relative humidity, when it becomes comparable with the value of $d_{(100)}$. In fit (4), pressure data are plot in a log-lin representation against the interlamellar distance, $d_{(100)}$, (Figure 3d) and then assume that $\Pi_{H1}$, $\Pi_{H2}$, $\lambda_{H1}$ and $\lambda_{H2}$ are simply extracted from a double linear fit according to equations Eq. 8 and Eq. 9. One should note that we have neglected the Van der Waals contribution, as in fit (2), and that in fit (4) only the slopes, providing $\lambda_{H1}$ and $\lambda_{H2}$ have a physical meaning, while the pressure values at the intercept, $\Pi_{H1}$ and $\Pi_{H2}$, do not.

**Table 1: Values of the hydration pressure ($\Pi_H$) and decay lengths ($\lambda_H$) in the primary and secondary hydration regimes. Data are obtained from the fit of the osmotic stress experiments in Figure 3a-d applying fits (1)-(4) to the low- (16 mM) and high-salt (100 mM) regimes. Assumptions: Fit (1) : parameters for $\Pi_{VdW}$ (Eq. 5): $H = 5.1 \cdot 10^{-21}$ J; $T_h$= 1.4 nm; $L$= 0.8 nm. Fit (2) : $\Pi(d_w) - \Pi_{VdW}$ with $\Pi_{VdW} \ll \Pi_{Hyd1}$; $\Pi_{Hyd2}$; Fit (3) : $\Pi(d_w) - \Pi_{VdW}$ with $\Pi_{VdW} \sim \Pi_{Hyd1}$; $\Pi_{Hyd2}$; Fit (4) : $\Pi(d_{(100)}) - \Pi_{VdW}$ with $\Pi_{VdW} \ll \Pi_{Hyd1}$; $\Pi_{Hyd2}$. This fit is performed against $d_{(100)}$. * Eq. 8 and Eq. 9 are used in their linearized form, as in Eq. 7.**

| Fit N° | Method | Equation | [Na$^+$] / mM | $\Pi_{H1}$/kbar | $\lambda_{H1}$/nm | $\Pi_{H2}$/kbar | $\lambda_{H2}$/nm |
|---|---|---|---|---|---|---|---|
| (1) | Fit (Lin-Lin) | Eq. 4 | 16 | 1.26·10$^3$ | 0.07 ± 20% | 1.81 | 2.98 ± 20% |
|  |  |  | 100 | 37.5 | 0.13 ± 10% | 1.61 | 1.59 ± 10% |
| (2) | Linear fit (Log-Lin) | Eq. 8* | 16 | 17.0 ± 40% | 0.28 ± 20% | 2.04 ± 15% | 2.53 ± 20% |
|  |  | Eq. 9* | 100 | 5.94 ± 7% | 0.45 ± 10% | 1.80 ± 12% | 1.43 ± 10% |
| (3) |  | Eq. 8* | 16 | 17.3 ± 40% | 0.28 ± 20% | 2.04 ± 15% | 2.50 ± 20% |

|   | Linear fit (Log-Lin) | Eq. 9* | 100 | 6.05 ± 7% | 0.45 ± 10% | 1.81 ± 2% | 1.42 ± 10% |
| --- | --- | --- | --- | --- | --- | --- | --- |
| (4) | Linear fit (Log-Lin) | Eq. 8* | 16 | 7.93·10$^6$ | 0.28 ± 20% | 8.47 | 2.52 ± 20% |
|  |  | Eq. 9* | 100 | 1.66·10$^4$ | 0.45 ± 10% | 8.47 | 1.42 ± 10% |
|  |  |  |  |  |  |  |  |
|  | **Average all** |  |  |  | **0.29 ± 0.15** |  | **2.05 ± 0.64** |
|  | **Average (Fit Log-lin)** |  |  |  | **0.37 ± 0.12** |  | **1.97 ± 0.78** |

Table 1 summarizes the ($\Pi_{H1}$, $\Pi_{H2}$, $\lambda_{H1}$, $\lambda_{H2}$) parameters obtained from the application of fits (1)-(4) on the pressure-distance curves obtained from the humidity chamber experiments performed on two GC18:0 (C = 1 wt%, pH = 6.3 ± 0.3) lamellar hydrogel samples at salt concentrations in the gel (prior to deposition onto the sample holder), [NaCl] = 16 mM and 100 mM. The following observations must be done:

a) *Agreement between our values and literature*. The values of $\lambda_{H1}$ and $\lambda_{H2}$, averaged over all fits, are respectively 0.29 ± 0.15 nm and 2.05 ± 0.64 nm. These values, despite the error (discussed here below) are characteristics for the short- and long-range decay lengths found in lamellar lipid systems characterized by primary and secondary hydration:[21,47] the values of the decay lengths are not dependent on the fit strategy.

b) *Impact of the fit*. Hydration forces are known to be very sensitive to salt concentration, and for this reason we run two experiments at [NaCl] = 16 mM and 100 mM. These values are the "bulk" values, and one should consider a five-fold increase in concentration in the humidity chamber, as already commented above. When using fits (2)-(4), $\lambda_{H1}$ and $\lambda_{H2}$ are highly homogeneous at each salt concentration, e.g., $\lambda_{H1}$= 0.28 nm and $\lambda_{H2}$= 2.52 nm at [NaCl] = 16 mM. On the contrary, fit (1) provides values of the decay length $\lambda_{H1}$, which are smaller by a factor three in the short range hydration respect to the values obtained using fits (2)-(4). Estimation of the longer decay lengths $\lambda_{H2}$ are also slightly different between (1) and (2)-(4), but still comparable within the error. The poor results of fit (1) are particularly visible in the values of the pressure, whereas fits (2)-(3) provide $\Pi_{H1}$ in the order of several kbar, while fit (1) provides an exceedingly high value of 10$^3$ kbar, which is not realistic. The above illustrates how the large error in the average values are mainly directed by the poor estimates obtained from fit (1).

c) *Impact of salt concentration*. Figure 3a-d shows that salt has little influence at short interlamellar distances (typically $d_w$< 0.7 nm), where the data at 16 mM and 100 mM are practically superimposed. Nonetheless, the limited number of experimental points recorded provide two distinct values of $\lambda_{H1}$, respectively 0.28 nm at 16 mM and 0.45 nm at 100 mM (analysis is here limited to fits (2)-(4) only). Nonetheless, these values are still comparable within the error, providing an average $\lambda_{H1}$= (0.37 ± 0.12) nm. This value and its small dependence on salt concentration are in strong agreement with what is found in primary hydration forces, generally related to enthalpic adsorption energy of water layers.[21,43,47] When it comes to secondary hydration at longer distances, Figure 3a-d show a strong impact of the initial salt concentration on the pressure-distance profiles. The corresponding decay length, $\lambda_{H2}$, are in worst agreement among themselves $\lambda_{H2}$= (1.97 ± 0.78) nm, with a relative error of about 40%, and they highlight the strong impact of salt. These aspects are in agreement with the literature data on secondary hydration,[10,13,21,22,43,47,48] of which the origin was attributed to the competition between water bound to the counterions and water bound to the bilayer surface.[43]

**Conclusion.**

We have used four fitting strategies to fit the pressure-distance curves of the GC18:0 lamellar phase. We have explored the short-range regime at water thicknesses below 3 nm. This regime is nicely probed by the humidity chamber apparatus at low and high salt content, which is estimated, in the dehydrated lamellar phase inside the humidity chamber, to be between 100 mM and 500 mM, respectively corresponding to [NaCl]= 16 mM and 100 mM in the bulk gel. The experimental data are fit using a double exponential decay, rendering primary and secondary hydration forces, the latter due to the hydration of the counterions. Whichever fit is used, the interdigitated layers in the sample experience two hydration regimes with decay lengths at about 0.3 nm and 2 nm, as expected from the literature on charged lipid lamellar phases containing salt. Both the choice of the fit, the amount of salt and the limited number of points generate an expected, although mild, dispersion in the hydration pressures and decay lengths. Altogether, the fitting strategies provide a consistent set of data for both the primary and secondary hydration regions.

**Acknowledgements**

We thank Dr. S. Roelants and Prof. Wim Soetaert (Gent University, Belgium) for producing the GC18:0 compound and Dr. E. Delbeke and Prof. C. Stevens (Gent University, Belgium) for the hydrogenation reaction. Access to the D16 beamline was financed by ILL under the proposal number 9-13-783, DOI: 10.5291/ILL-DATA.9-13-783. We thank Dr. Bruno Demé for his kind assistance on the experiments.